\documentclass[useAMS,usenatbib]{mn2e}
\usepackage{graphicx}
\usepackage{amssymb}
\usepackage{amsmath}
\title{Constraining Primordial Black Hole Fraction at the Galactic Centre using radio observational data}
\author[Chan \& Lee]{Man Ho Chan \thanks{chanmh@eduhk.hk}, Chak Man Lee
\\ Department of Science and Environmental Studies, The Education University of Hong Kong, Tai Po, Hong Kong}

\begin{document}

\date{Accepted XXXX, Received XXXX}

\pagerange{\pageref{firstpage}--\pageref{lastpage}} \pubyear{XXXX}

\maketitle

\label{firstpage}

\date{\today}

\begin{abstract}
Recent gamma-ray and cosmic-ray observations have put strong constraints on the amount of primordial black holes (PBHs) in our universe. In this article, we use the archival radio data of the inner Galactic Centre to constrain the PBH to dark matter ratio for three different PBH mass distributions including monochromatic, log-normal and power-law. We show that the amount of PBHs only constitutes a very minor component of dark matter at the Galactic Centre for a large parameter space.
\end{abstract}

\begin{keywords}
Primordial black holes
\end{keywords}

\section{Introduction}

Primordial black hole (PBH) has long been of considerable interests for nearly half a century \citep{Zel'dovich, Khlopov}, since merely it could be sufficiently small for Hawking radiation \citep{Hawking, Hawking1} among all types of black holes. Such a hypothetical type of massive compact halo object is naturally regarded as a good dark matter candidate because it does not involve any new particles beyond the Standard Model. Also, PBHs are nearly collisionless and they are stable if sufficiently massive. If they still exist in a large number by now, they would contribute to a large fraction of dark matter, and Hawking radiation from them may be an appreciable source of observable background radiation.

These massive compact objects would be generated in the early Universe less than one second just after the Big Bang. In general, their expected masses depend on the time, at which they were created, ranging from the Planck mass ($10^5$ g) created at the Planck time ($t\approx 10^{-43}$ s) to about $10^5 M_{\odot}$ just before big-bang nucleosynthesis ($t\approx 1$ s). In other words, such realistic production mechanisms expect an extended mass distribution rather than just a unique mass for all PBHs.

Owing to such a very large mass range, the past decades of various observations, including intensive 511 keV gamma-ray line at the Galactic Centre \citep{Prantzos, Siegert} and recent detection of gravitational waves from binary black hole merger \citep{Abbott, Abbott1,Bird}, revive a great deal of interests on exploring mass windows or further improving constraints on the PBH to dark matter ratio \citep{Carr,Bernard,Clesse,Riccardo,Gnedin,Zhu,Boudaud,Green,Bellomo,Ranjan,DeRocco,Alexandre,Paulo,Chan2,Katza}. For astrophysical and cosmological observations, \citet{Clesse} give seven hints in favor of a PBH population with abundance comparable to dark matter, coming from various observations of widely different scales and epochs of the Universe or revealed by using physical arguments based on first principles. Among those mass constraints, sufficiently small masses ($<4 \times 10^{14}$ g) are ruled out since such PBHs are believed to evaporate away completely through Hawking radiation \citep{Hawking, Hawking1} much earlier. For PBH masses $\ge 4 \times 10^{14}$ g, various kinds of data such as the observational data of the Voyager 1 Satellite \citep{Stone, Cummings, Boudaud}, strong lensing measurements \citep{Paulo, Katza}, background gamma-ray and cosmic-ray data \citep{Carr2,Ranjan,Carr3} have been considered to constrain the fraction of PBH being dark matter $f$ (the mass ratio of total PBHs to total dark matter). For some mass windows, most of these data show that $f$ is much less than 1, which means that PBHs may only constitute a very small part of dark matter \citep{Carr2,Ranjan,Boudaud,Carr3}. Nevertheless, some mass ranges of PBHs are still viable for $f=1$ \citep{Smyth,Dasgupta,Laha2}. 

On the other hand, some recent studies focus on different classes of mass distributions of PBHs \citep{Boudaud, Green, Bellomo, Ranjan} and even PBHs of Kerr type or spinning PBHs \citep{Alexandre,Kuhnel,Dasgupta}. They all conclude that the constraints for PBH with a wide range of PBH mass are more stringent than that with a unique mass only. 

Apart from using gamma-ray or cosmic-ray data to constrain PBH, in this article, we explore the possibility of using radio data to constrain PBH fraction. We also examine the PBH fraction at the inner Galactic Centre, which has not been widely discussed before \citep{Carr3}. Therefore, the objective of the present article is to update the constraints on PBH fraction at the inner Galactic Centre by radio observational data. Similar analyses have been done in our previous work to constrain dark matter annihilation model \citep{Chan, Chan1}. The high energy particles produced from the evaporation of PBHs would emit synchrotron radiation in radio bands when there is a strong magnetic field. By using the radio data of the Galactic Centre, we can obtain the upper limits of the ratio of PBH mass density to dark matter density. These results are generally consistent with that obtained by gamma-ray and cosmic-ray observations.

\section{Formalism for the PBH evaporation model}
Similar to traditional black holes, PBHs with mass $M_{\rm PBH}$ would emit Hawking radiation and particles (i.e. PBH evaporation). The temperature as a unique parameter linking with their mass $M_{\rm PBH}$ dictates the rate of the particles emitted from the PBHs' surface and is given by \citep{Hawking,Hawking1}
\begin{equation}
k_{\rm B}T=\frac{\hbar c^3}{8\pi GM_{\rm PBH}}\approx 1.06\left(\frac{10^{13}{\rm g}}{M_{\rm PBH}}\right) {\rm GeV}.
\end{equation}
The number of particles $N_e$, especially electron-positron pairs, emitted per unit time and energy follows the distribution \citep{Boudaud}
\begin{equation}
\frac{d^2N_e}{dEdt}=\frac{\Gamma_e}{2\pi}\left[\exp\left(E/k_{\rm B}T\right)+1\right]^{-1},
\end{equation}
where the electron absorption probability $\Gamma_e$ can be approximately modelled as \citep{MacGibbon}
\begin{equation}
\Gamma_e\approx\left\{
\begin{array}{ll}
16 \frac{G^2M_{\rm PBH}^2E^2}{\hbar^3c^6}\,\,\,\,{\rm for}\,\,\,\, \frac{GM_{\rm PBH}E}{\hbar c^3}< 1,\\
27 \frac{G^2M_{\rm PBH}^2E^2}{\hbar^3c^6}\,\,\,\,{\rm for}\,\,\,\, \frac{GM_{\rm PBH}E}{\hbar c^3}\ge 1.
\end{array}\right.
\end{equation}
Beside the primary emission of electron-positron pairs, PBH evaporation would also give quark pairs which would quickly hadronize to produce secondary emission of electrons and positrons. Nevertheless, the energy of the secondary electrons and positrons is somewhat lower than the primary ones. The spectrum of the secondary electron-positron pairs produced can be calculated by the BlackHawk code published in \citet{Arbey}. We show the spectra of the primary and secondary emissions from PBH evaporation for $M_{\rm PBH}=10^{14}$ g in Fig.~1.

In the followings, we consider two classes of mass distribution of PBH: monochromatic mass distribution and extended mass distribution. Starting from the former case, we assume that all PBHs have a common mass following a monochromatic mass distribution. The number of injected electrons per unit time, energy and volume is given by
\begin{equation}
Q(E,r)=\frac{\rho_{\rm PBH}(r)}{M_{\rm PBH}}\frac{d^2N_e}{dEdt},
\label{monochromatic}
\end{equation}
where $\rho_{\rm PBH}(r)$ denotes the total PBH mass density, which is a function of the distance from the Galactic Centre, $r$. Most of the previous studies assume that the total PBH mass density traces dark matter density \citep{Carr}. It is because dark matter density usually dominates the mass density in stable structures so that PBH density distribution follows dark matter density distribution via the gravitational interaction between PBHs and dark matter. Therefore, the PBH fraction is usually defined as $f=\rho_{\rm PBH}/\rho_{DM}$, where $\rho_{DM}$ is the dark matter density. We follow this definition and assume that the PBH distribution traces the dark matter distribution. Based on this definition, we have
\begin{equation}
Q(E,r)=\frac{f \rho_{DM}(r)}{M_{\rm PBH}}\frac{d^2N_e}{dEdt}.
\label{fraction}
\end{equation}

We also generalize the case of extended mass distribution for a more realistic situation. The number of injected electrons per unit time, energy and volume can be written as
\begin{equation}
Q(E,r)=\frac{f \rho_{DM}(r)}{\rho_{\odot}}\int_{\forall M_{\rm PBH} } dM_{\rm PBH}\frac{g(M_{\rm PBH})}{M_{\rm PBH}}\frac{d^2N_e}{dEdt},
\label{extended}
\end{equation}
where $g(M_{\rm PBH})$ is any particular mass distribution of PBH normalized to $\rho_{\odot}$. $\int_{\forall M_{\rm PBH} } dM_{PBH}$ indicates the integration over a large PBH mass range which includes our interested range $4 \times 10^{14} {\rm g}\le M_{\rm PBH}\le 10^{17} {\rm g}$. In the current study, we focus on two different popular extended mass distributions: log-normal distribution and power-law distribution.

The log-normal distribution \citep{Boudaud,Green,Ranjan,Bellomo,Alexandre}, with $\mu$ and $\sigma$ denoted as the median and the standard derivation of the logarithm of the mass distribution respectively \citep{Krishnamoorthy}, is defined as
\begin{equation}
g(M_{\rm PBH})=\frac{\rho_{\odot}}{\sqrt{2\pi}\sigma M_{\rm PBH}}\exp\left(-\frac{\ln^2(M_{\rm PBH}/\mu)}{2\sigma^2}\right),
\end{equation}
whereas the power-law distribution, as parameterized by the power-law index, $p$, the maximum value $M_{\rm max}$, and the minimum value $M_{\rm min}$ of the mass distribution, is defined as \citep{Ranjan,Bellomo},
\begin{equation}
g(M_{\rm PBH})=\frac{p \rho_{\odot}}{M_{\rm max}^{p}-M_{\rm min}^{p}}M_{\rm PBH}^{p-1},
\end{equation}
where $M_{\rm PBH}\in [M_{\rm min},M_{\rm max}]$ and the power-law index $p	\neq 0$.

In the followings, we will focus on the central region ($r \le 0.16$ pc) of the Galactic Centre. If there exists a large amount of PBHs, the evaporation of PBHs would give a large amount of high-energy electrons and positrons ($\sim 0.001-0.1$ GeV). These high-energy electrons and positrons would emit synchrotron radiations when they are subjected in a magnetic field strength $B$ and surrounded by a background plasma with plasma frequency $\nu_p = 8890 [n(r)/1\,\, {\rm cm}^{-3}]^{1/2}$ Hz, where the electron density is $n(r)\sim 1~{\rm cm}^{-3}$ \citep{Muno}. The resulting synchrotron power is very insensitive to the electron density $n(r)$ (the change is negligible for $n(r)=0.1-10$ cm$^{-3}$), and the synchrotron radiations can be easily detected by radio telescopes. The average synchrotron power for a single electron with energy $E=\gamma m_ec^2$ ($\gamma$ is the Lorentz factor of the electron), at a particular frequency $\nu$, can be expressed as \citep{Storm}
\begin{equation}
P_{\rm syn}=\int_0^{\pi}d\theta\frac{\sin^2\theta}{2}2\pi\sqrt{3}r_e m_e c \nu_g F_{\rm syn}\left(\frac{x}{\sin\theta}\right),
\label{syn}
\end{equation}
where $\nu_g=eB/(2\pi m_ec)$ is the non-relativistic gyrofrequency, $r_e=e^2/(m_ec^2)$ is the classical electron radius and $\theta$ is the pitch angle. The quantities $x$ and $F$ are defined as
\begin{equation}
x=\frac{2\nu}{3\nu_g\gamma^2}\left[1+\left(\frac{\gamma\nu_p}{\nu}\right)^2\right]^{3/2},
\end{equation}
and
\begin{equation}
F_{\rm syn}(y)=y\int_y^{\infty}K_{5/3}(s)ds \approx 1.25y^{1/3}e^{-y}(648+y^2)^{1/12},
\end{equation}
respectively. When high-energy electrons and positrons are produced, they would diffuse and cool down via the diffusion-loss equation \citep{Atoyan, Chan, Chan1}. Due to a very large magnetic field strength at the Galactic Centre ($B \sim 1$ mG), the cooling of electrons and positrons would be dominated by the synchrotron cooling. The cooling rate (in the unit of $10^{-12}$ GeV s$^{-1}$) is given by \citep{Colafrancesco}
\begin{equation}
b(E)=2.54E^2B^2,
\end{equation}
where $E$ and $B$ are in the units of GeV and mG, respectively. Note that the above cooling rate is applicable for ultra-relativistic electrons and positrons only. Nevertheless, the cooling of semi-relativistic ($E<0.01$ GeV) and non-relativistic electrons and positrons (mainly the secondary emission due to hadronization) would be dominated by other processes such as Coulomb cooling and Bremsstrahlung cooling. The synchrotron emission due to these electrons and positrons is negligible. Later, we will see that our results are mainly based on the contribution of the primary emission (ultra-relativistic electrons and positrons). Therefore, using the cooling rate in Eq.~(12) is a very good approximation. The estimated cooling time for a 0.01 GeV electron is $t_c \sim 10^{13}$ s. For a very small region with a very high magnetic field strength, the stopping distance of a positron or electron based on the simple random walk model is $d_s \sim \sqrt{r_Lct_c}$, where $r_L=E/ecB$ is the Larmor radius. The stopping distance for a 0.01 GeV electron is about $10^{-3}$ pc, which is smaller than the size of the region we considered (radius $R=0.16$ pc) at the Galactic Centre. Also, the advection is important only for $r \le 0.04$ pc \citep{Bertone}. Therefore, most of the electrons and positrons produced by the evaporation of PBHs in this small region would be almost confined in that region. The solution for the equilibrium state can be expressed, in terms of $Q(E,r)$, by
\begin{equation}
\frac{dn_e}{dE}=\frac{1}{b(E)}\int_E^{\infty} Q(E',r)dE',
\end{equation}

By assuming that the PBH distribution is spherically symmetric, the radio flux density emitted due to the evaporation of the PBHs, with sufficiently large distance $D_L=8.5$ kpc to the region considered at the Galactic Centre as observed from the Earth, is finally given by \citep{Chan,Chan1}
\begin{equation}
S_{DM}(\nu)=\frac{1}{4\pi D_L^2}\int_0^R\int_{m_e}^{\infty}2\frac{dn_e}{dE}P_{\rm syn}dE(4\pi r^2)dr,
\end{equation}
which depends on the radio frequency $\nu$. The factor 2 in the above equation indicates the contributions of both high-energy electrons and positrons.

\section{Results}
We examine three possible dark matter density profiles at the Galactic Centre for $r \le 0.16$ pc: the Navarro-Frenk-White (NFW) density profile $\rho_{DM}=\rho_sr_s/r$ with $(\rho_s,r_s)=(0.0182M_{\odot}{\rm pc^{-3}},10.7~{\rm kpc})$ \citep{Sofue}, a cored profile (or the cored isothermal profile) $\rho_{DM} \approx \rho_s=0.0365M_{\odot}{\rm pc^{-3}}$ \citep{Cirelli} and a contracted density profile $\rho_{DM}=\rho_s(r/r_s)^{-1.5}$ with $(\rho_s,r_s)=(100M_{\odot}{\rm pc^{-3}},2~{\rm pc})$ \citep{Gnedin}. For the magnetic field strength, its profile near the Galactic Centre can be modeled by the following form \citep{Bringmann}
\begin{equation}
B=\left\{
\begin{array}{ll}
B_0 \left(\frac{r_c}{r}\right)^{5/4}& {\rm for }\,\,\, r\le r_c,\\
B_0 \left(\frac{r_c}{r}\right)^2& {\rm for }\,\,\, r > r_c.\\
\end{array}
\right.
\end{equation}
where the magnetic field at radius $r_c=0.04$ pc is $B_0= 7.2~{\rm mG}$.

In Fig.~2, we show the radio flux spectra for two different $M_{\rm PBH}$ based on the monochromatic mass distribution model (assumed the NFW profile with $f=0.01$). The resulting radio flux density depends largely on the PBH mass and it can range from $10^{-8}$ Jy to $10^4$ Jy. Moreover, synchrotron emission is proportional to the energy of electrons and positrons. As mentioned above, since the energy of the secondary electrons and positrons produced from PBH evaporation dominates in the low energy range ($\sim 0.001-0.01$ GeV) (see Fig.~1), the radio emissions due to the secondary emission of electrons and positrons are nearly negligible (see Fig.~2). Therefore, the primary emission spectrum in Eq.~(2) can almost represent the contribution of radio emission. 

The archival data in \citet{Davies} reveal that the upper radio flux density limit within a radius of 4" (0.16 pc) at the Galactic Centre at frequency 0.408 GHz is 50 mJy. This information can give a stringent constraint on the PBH fraction $f$. In Fig.~3, we show the upper limits of $f$ as a function of PBH mass for the monochromatic mass distribution for three different dark matter density profiles. The region in between the limits assuming the cored profile and the contracted profile indicates the systematic uncertainty band due to the dark matter density. These limits are consistent with the previous studies using X-ray, gamma-ray and cosmic-ray data \citep{Ranjan,Ballesteros,Boudaud,Laha2}. We also plot the upper limits of $f$ for the log-normal distribution and the power-law distribution in Fig.~4 and Fig.~5 respectively. Certain ranges of parameters (e.g. $\sigma$, $p$) have been assumed in the plots. These limits are generally tighter than that using gamma rays and cosmic rays. For a considerable size of the parameter space, the value of $f$ is much less than 1, which means that the amount of PBHs is quite insignificant at the inner Galactic Centre.

\begin{figure}
\vskip 3mm
\includegraphics[width=80mm]{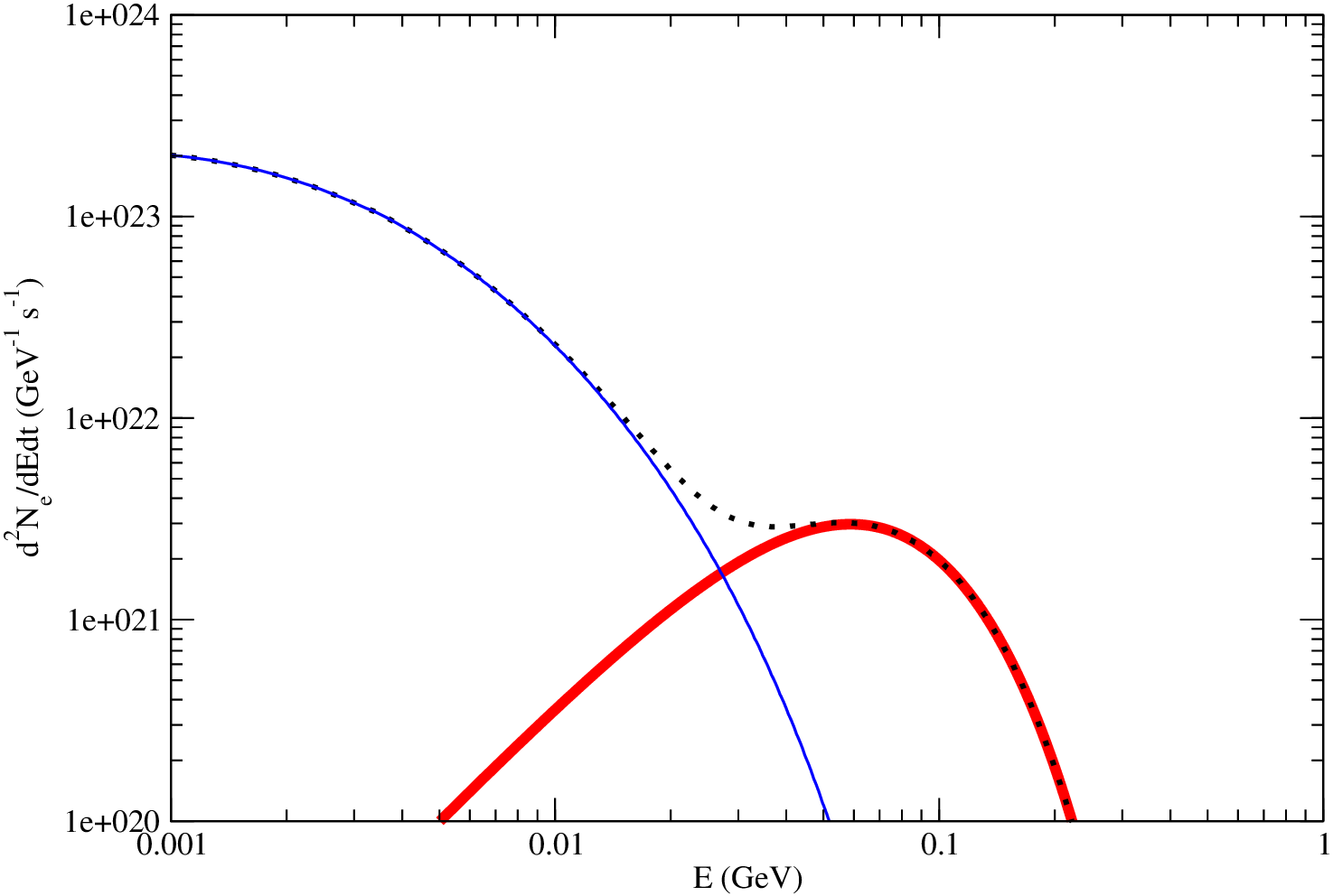}
\caption{The spectra of primary emission (red line) and secondary emission (blue line) of electrons or positrons for $M_{\rm PBH}=10^{14}$ g. The black dotted line is the total spectrum of electron or positron emission (primary plus secondary emission).}
\label{Fig1}
\end{figure}

\begin{figure}
\vskip 3mm
\includegraphics[width=80mm]{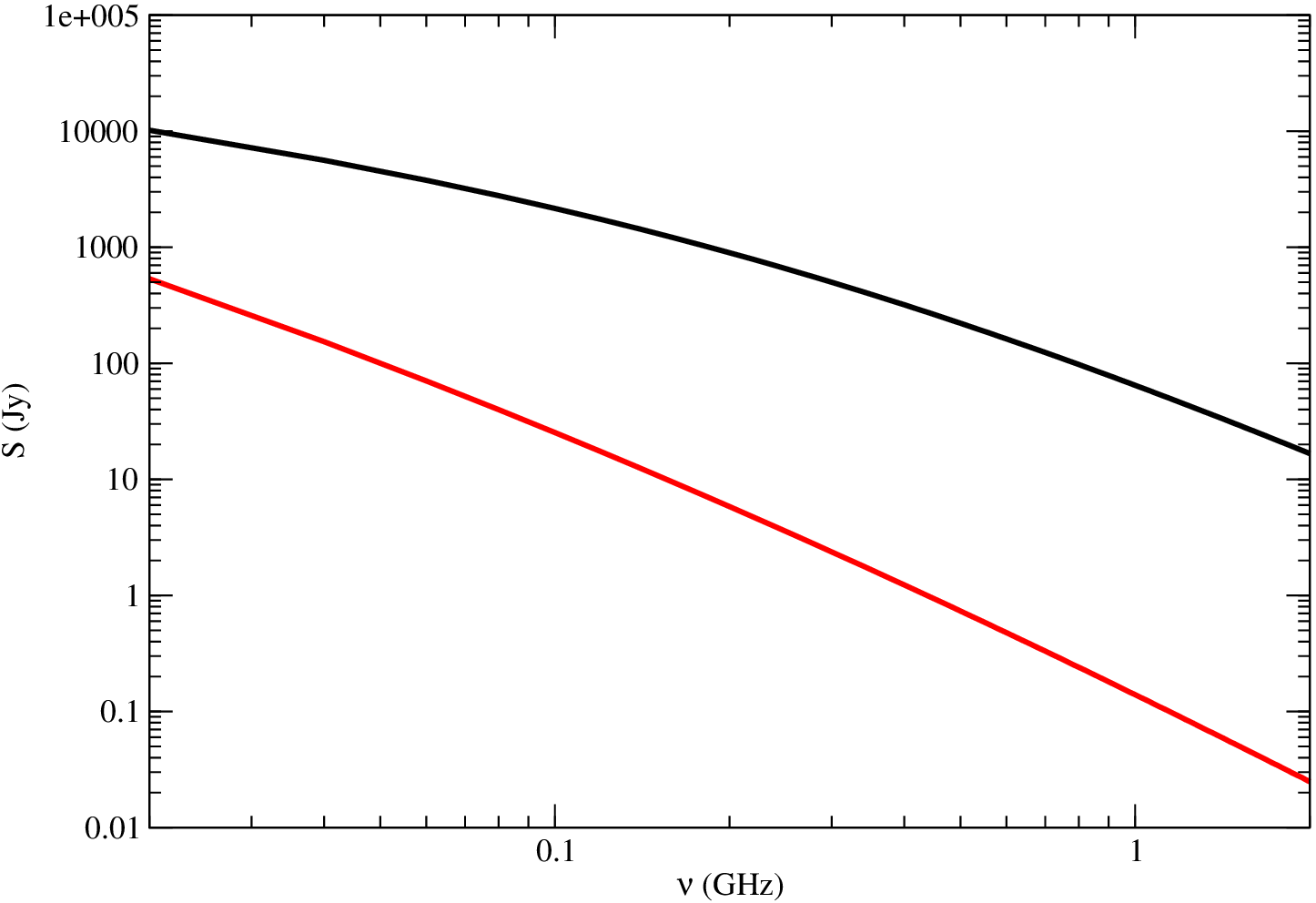}
\vskip 10mm
\includegraphics[width=80mm]{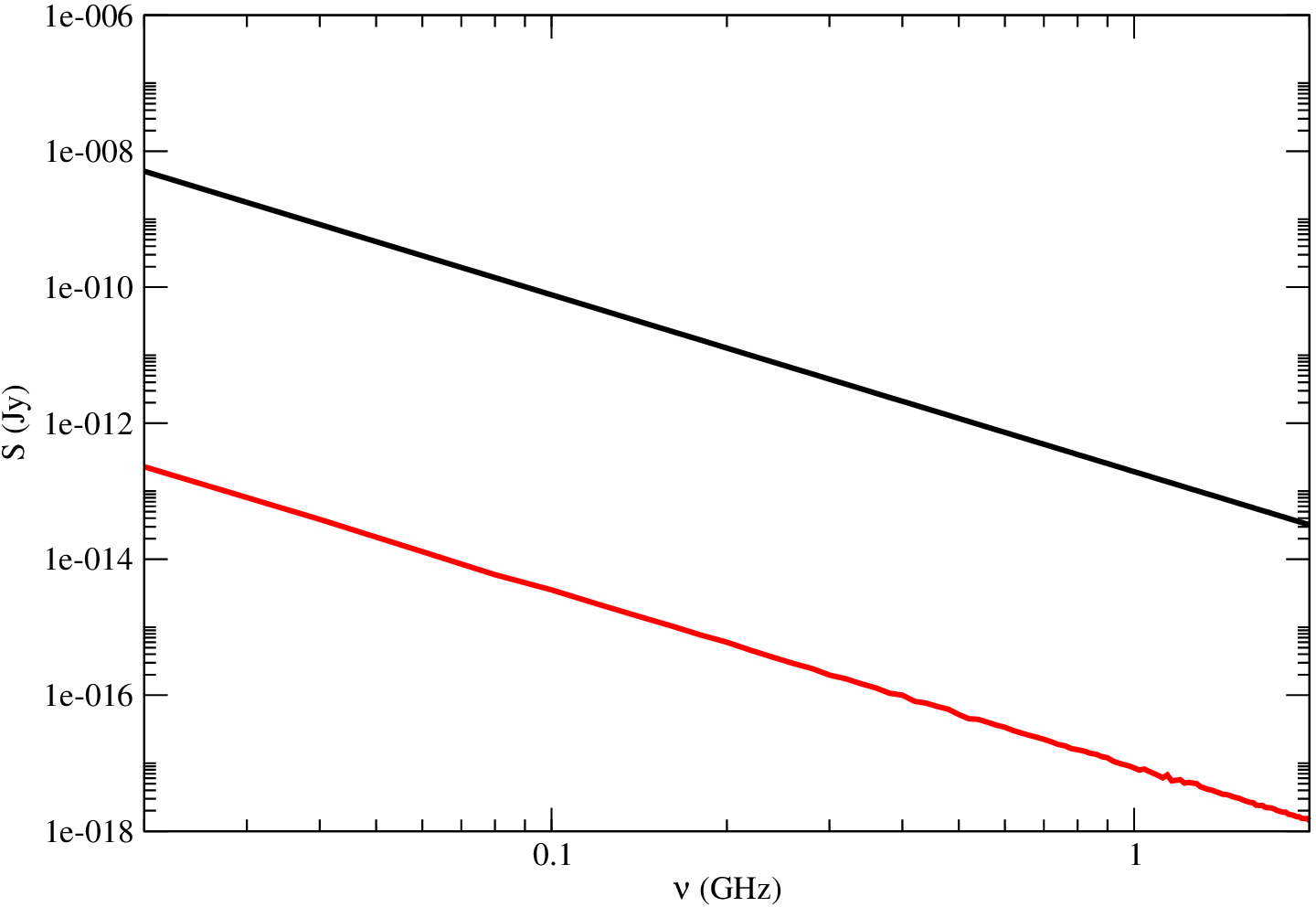}
\caption{The predicted radio flux densities from PBHs as a function of frequency originated from primary emission (black lines) and secondary emission (red lines), with PBH monochromatic masses $M_{\rm PBH}=4 \times 10^{14}~{\rm g}$ (upper) and $M_{\rm PBH}=10^{17}~{\rm g}$ (lower), assuming the NFW profile with $f=0.01$.}
\label{Fig2}
\end{figure}

\begin{figure}
\vskip 3mm
\includegraphics[width=80mm]{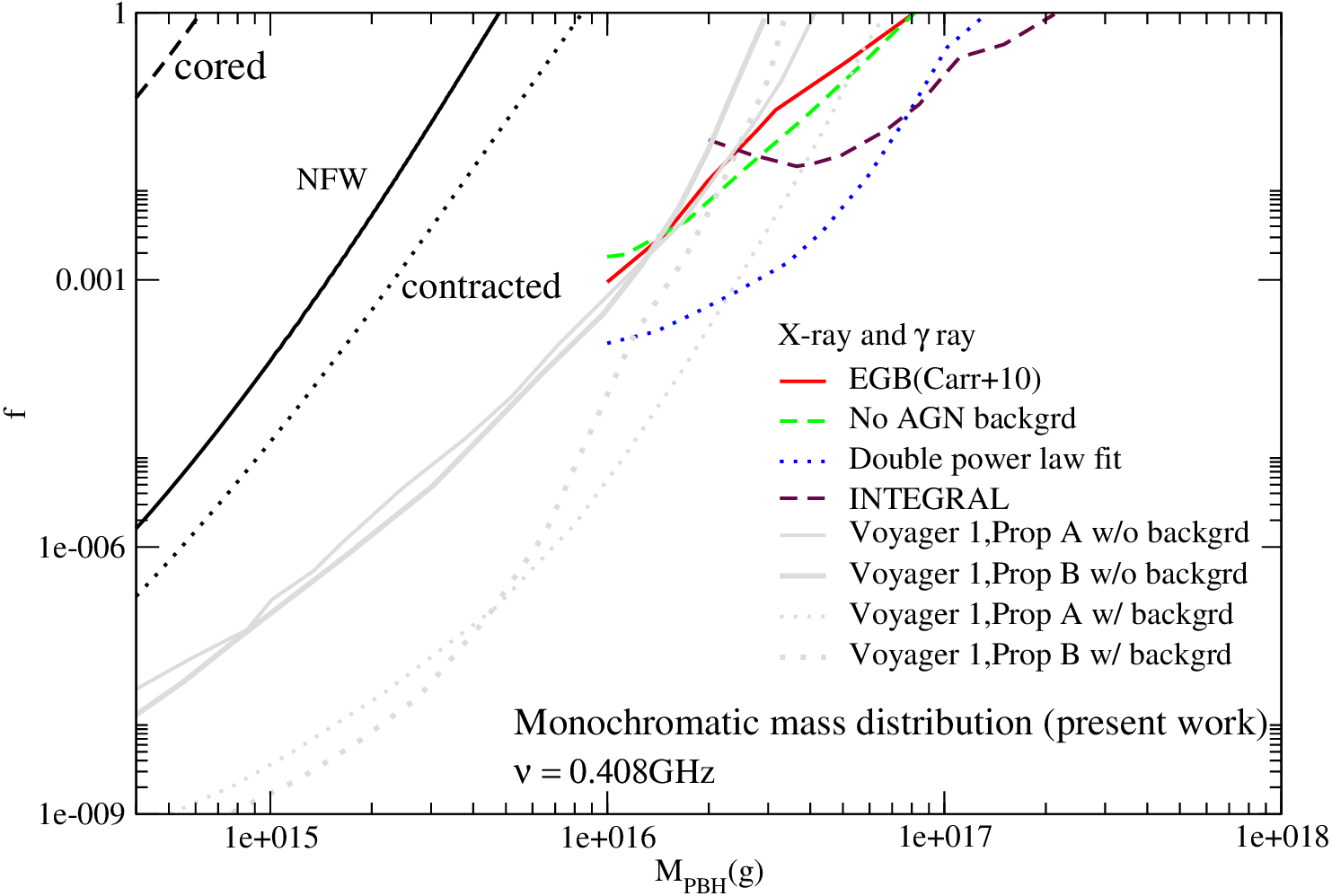}
\caption{The upper limits of $f$ as a function of the PBH monochromatic mass. The black solid line, dashed line and dotted line represent the limits in this analysis for the NFW profile, the cored profile and the contracted profile respectively. The coloured lines indicate the upper limits of $f$ constrained by the cosmic-ray, gamma-ray and X-ray data \citep{Boudaud,Ballesteros,Laha2}.}
\label{Fig3}
\end{figure}

\begin{figure}
\vskip 3mm
\includegraphics[width=80mm]{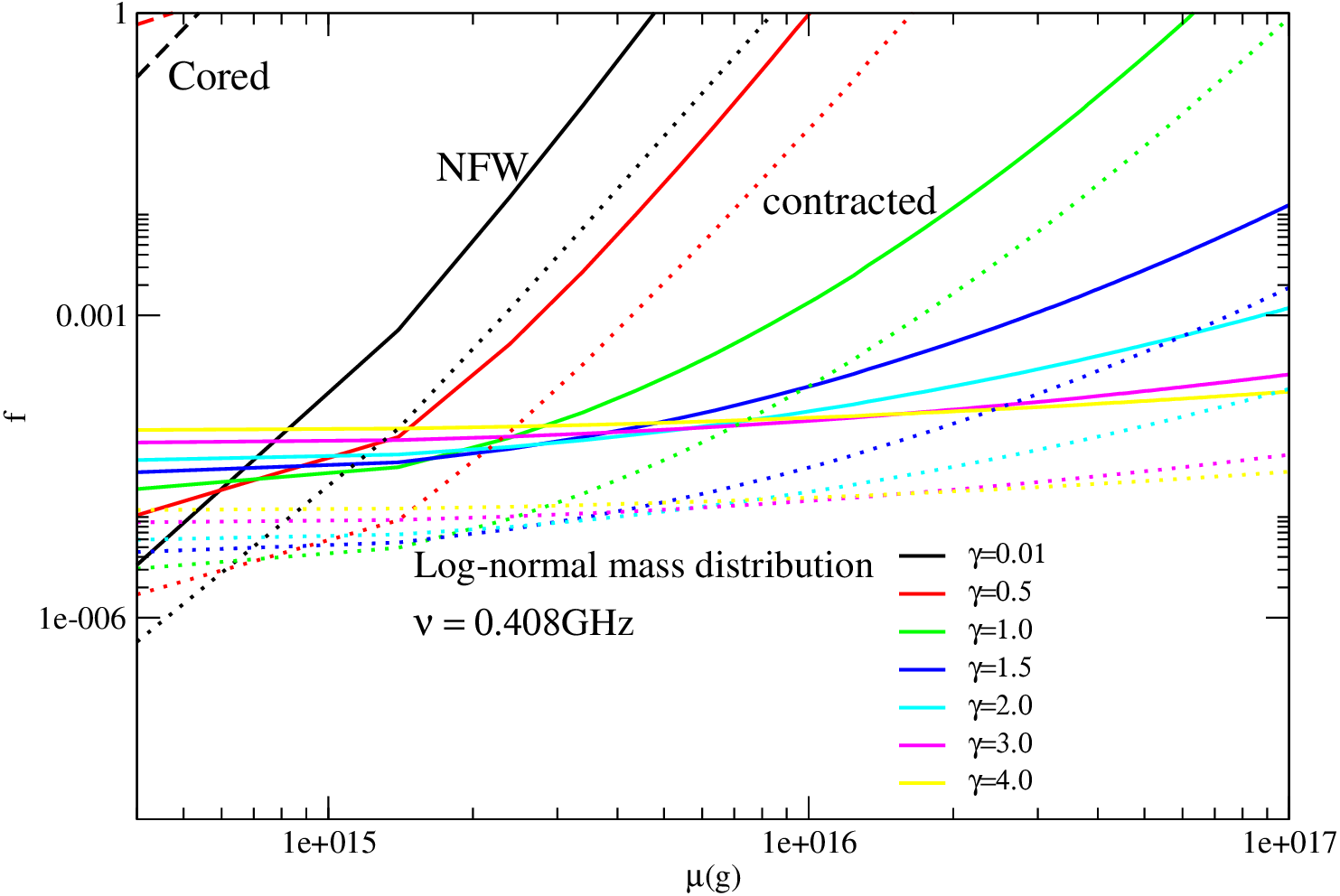}
\caption{The upper limits of $f$ as a function of the median PBH mass $\mu$, assuming the log-normal mass distribution.}
\label{Fig4}
\end{figure}

\begin{figure}
\vskip 3mm
\includegraphics[width=80mm]{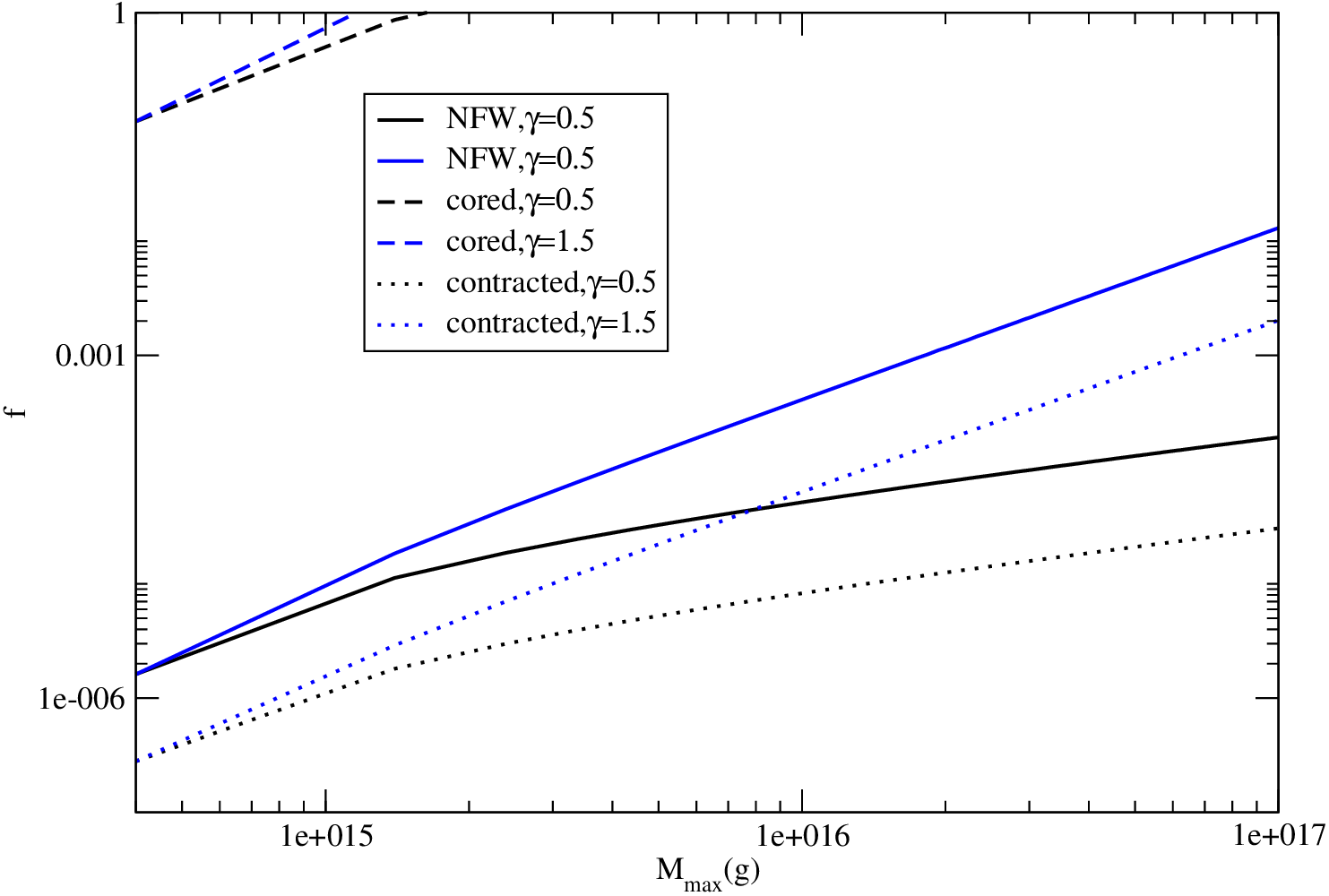}
\includegraphics[width=80mm]{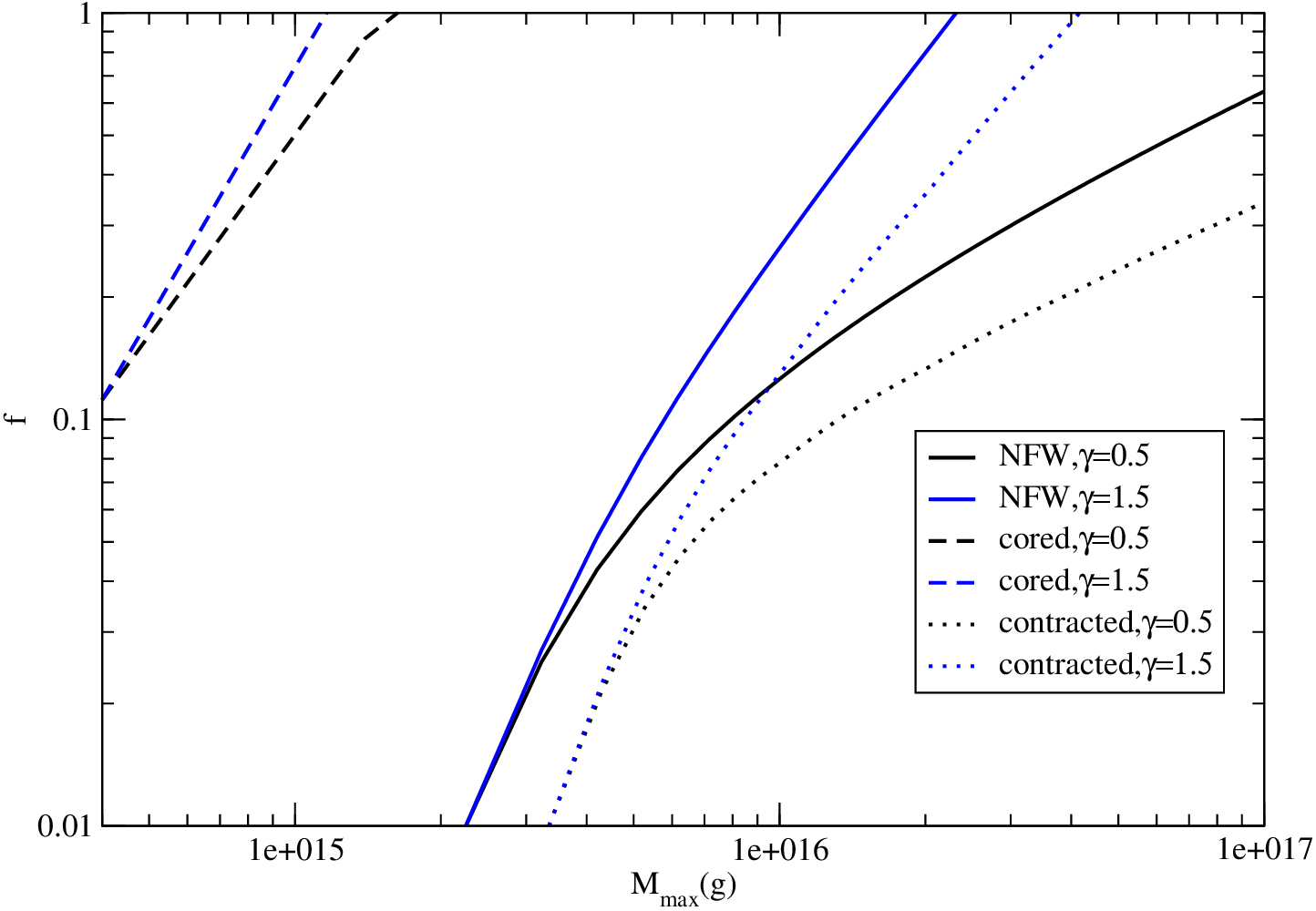}
\caption{The upper limits of $f$ as a function of the maximum mass $M_{\rm max}$, assuming the power-law mass distribution with two different power-law index $p=0.5$ and 1.5 for comparison. Here, the two PBH minimum masses assumed are (a) $M_{\rm min}=4\times 10^{14}$ g and (b) $M_{\rm min}=7.9\times 10^{15}$ g.}
\label{Fig5}
\end{figure}

\section{Discussion}
In the above studies, we constrain the PBH fraction at the Galactic Centre using the archival radio data. The advantage of using radio data is that radio flux density can be obtained with a very high resolution and sensitivity. For example, the data used in this analysis were collected within a radius of 4" near the Galactic Centre. This resolution is much greater than that obtained by gamma-ray observations. The main reason for considering the small region at the Galactic Centre is that it has a very high magnetic field strength ($B \sim 5$ mG), which can give a considerable amount of radio flux in current observable frequencies ($\nu \ge 0.1$ GHz). The synchrotron peak frequency (the radio frequency where the synchrotron emission is maximum) is $\nu \approx 4.7~{\rm GHz}(E/{\rm GeV})^2(B/{\rm mG})$ \citep{Profumo}. For $B \sim 5$ mG and $E \sim 0.1$ GeV, the peak frequency is 0.24 GHz, which is still close to the current observable radio frequency range. For other regions in our Galaxy, the magnetic field strength is about $B \sim 1-10$ $\mu$G so that the peak frequency would be smaller than $10^{-3}$ GHz, which is far below the current observable radio frequency range. Therefore, to constrain the PBH fraction using radio data, we can only consider the central region of a galaxy, which has sufficiently large magnetic field strength.

Based on our analysis, we show that only a small parameter space is allowed for $f$ being close to 1 (except for the cored dark matter density profile). This means that the amount of PBHs is probably a very minor component at the inner Galactic Centre. This is consistent with the previous studies using gamma rays and cosmic rays \citep{Ranjan}. Note that some systematic uncertainties may be included in our analysis. The major systematic uncertainty is the functional form of the magnetic field strength near the Galactic Centre. The profile in Eq.~(15) is derived from an approximate equipartition of magnetic, kinetic and gravitational potential energy inside the accretion zone for $r<0.04$ pc and magnetic flux conservation for $r \ge 0.04$ pc \citep{Bringmann}. Although the order of magnitude of magnetic field strength based on this assumption is supported by recent observations \citep{Guenduez}, the actual variation of magnetic field strength near the Galactic Centre is quite complicated. Our simplified profile would contribute some systematic uncertainties in the analysis. 

Besides, the systematic uncertainty of the dark matter density profile is another possible uncertainty. In this analysis, we have considered three dark matter density profiles, the NFW profile, constant density cored profile and the contracted profile. These profiles basically represent a large range of possible dark matter density, from a conservative value (constant cored profile) to an optimistic value (contracted profile). The regions in between the limits assuming the cored profile and the contracted profile in Figs.~3-5 represent the involved systematic uncertainty bands. Nevertheless, many previous and recent studies have realised that the supermassive black hole at the Galactic Centre might have induced a contracted dark matter density profile \citep{Gondolo,Merritt,Gnedin,Fields}. Therefore, the limits derived from the contracted dark matter density profile may be more representative and realistic. 

\section{Acknowledgements}
This work was supported by a grant from the Research Grants Council of the Hong Kong Special Administrative Region, China (Project No. EdUHK 28300518).

\section{Data availability statement}
The data underlying this article will be shared on reasonable request to the corresponding author.

\label{lastpage}

\end{document}